\documentclass[epj]{svjour}
\usepackage{graphics}
\usepackage{subfigure}
\usepackage{url}
\usepackage[english]{babel}
\usepackage[utf8]{inputenc}

\begin{document}

\title{A Schelling model with switching agents:\\ decreasing segregation via random allocation and social mobility}
\titlerunning{A Schelling model with switching agents}
\author{Aur\'elien Hazan\inst{1} \and Julien Randon-Furling\inst{2}
\thanks{Both authors contributed equally to this work.}
}                     
%
%
\institute{LISSI, Universit\'e Paris-Est Cr\'eteil (UPEC), 36-37 rue Georges Charpak, 77567 Lieusaint, France \and SAMM, Universit\'e Paris-1 Panth\'eon-Sorbonne, Centre Pierre Mendès-France, 90 rue de Tolbiac, 75013 Paris, France}
\date{Received: date / Revised version: date}
%
\abstract{
We study the behaviour of a Schelling-class system in which a fraction~$f$ of spatially-fixed switching agents is introduced. This new model allows for multiple interpretations, including: (i) random, non-preferential allocation (\textit{e.g.} by housing associations) of given, fixed sites in an open residential system, and (ii) superimposition of social and spatial mobility in a closed residential system.\\
We find that the presence of switching agents in a segregative Schelling-type dynamics can lead to the emergence of intermediate patterns (\textit{e.g.} mixture of patches, fuzzy interfaces) as the ones described in~\cite{HaBe}. We also investigate different transitions between segregated and mixed phases both at $f=0$ and along lines of increasing $f$, where the nature of the transition changes.
} 
\maketitle
\section{Introduction}
\label{sec: intro}
\begin{sloppypar}
Since Schelling's seminal work \cite{Schel1} introducing stochastic modeling on a checkerboard to gain insights about segregation phenomena in urban environments, much effort has been devoted to develop and understand multi-agent systems subject to Schelling dynamics. The existence of similarities with certain types of spin-system and liquid-solid dynamics \cite{MO,VKir,StauSol,MSSt,SRS,GVN09,GVNth} has been instrumental both in drawing theoretical physicists' attention to segregation phenomena and in providing the basis for advanced analogies with existing results and methods from the statistical physics toolbox. Many variants of Schelling's initial dynamics have been found to exhibit similar properties, and the main characteristic, namely the emergence of segregation, was shown recently to be a universal property common to a vast class of systems \cite{RM11,Grau2}.

We investigate in this article the impact of introducing, within a Schelling-class dynamics with two types of agents, a given fraction $f$ of fixed agents able to switch, that is: to change spontaneously from one of the two types of agents to the other, rather than simply bearing the same type throughout the simulation. This amounts to imposing a noise or a perturbation in the form of a spatially-fixed random background within the Schelling dynamics and we depart here from the standard context where agent types usually correspond to ethnic groups that cannot change. Let us emphasize that the dynamics studied here is \textit{a priori} different from Glauber and Kawasaki dynamics as discussed in \cite{StauSol}, since we follow Schelling and impose a non-zero vacancy density and do not allow for direct site-exchange between agents. Our model can thus be viewed as an interpolation between a (trivial) dilute system of non interacting particles in zero external field (at $f=1$, all agents are fixed) and a magnetic system of moving and interacting particles with zero total magnetization (at $f=0$, agents follow preferential-choice dynamics and there is an equal number of agents of both types).\footnote{The authors wish to thank an anonymous referee for drawing their attention to this particular point.}

Our motivation is twofold: on the theoretical side, the complexification introduced here differs from changes in the agents' interaction energies (or utility functions in socio-economic contexts) that have been explored in the literature, and we would like to see how it modifies the system's phase diagram. On a more heuristic side, the question is whether the presence of switching agents tends to facilitate mixing and de-segregation, and the aim is to explore a new variant of residential dynamics, with a long-term view to identifying factors that have the logical, mathematical capacity to lead to desegregation or lesser segregation, just as ``social distance and preference dynamics do have the logical capacity to combine with socioeconomic inequality between groups to create relatively high levels of ethnic and class segregation'' \cite{ClFos}.
\medskip

The new specification we suggest may be interpreted in at least two ways. First, the agents able to switch, being spatially fixed, can be viewed as housing sites that are never empty and for which the landlords (eg government authorities or housing associations) enforce an allocation policy that is blind to agents' types. In such a case, the system should be considered as ``open'' in the sense that when an agent switches types, it is as if an agent of the former type had moved outside of the system and an agent of the new type had moved in (the total number of agents in the system is fixed, but not that of agents of a given type). Secondly, if we favour a ``closed''-system view, switching can be interpreted as ``social mobility''. The easiest may be to picture agent types as social groups defined \textit{eg} by income levels. In this case, there can be various reasons why agents switching from a social group to the other may remain fixed spatially. Indeed, an agent that switches from the higher-status type to the lower one may wish to cling to her former status, as embodied by the dominating type in her neighbourhood. If the switching occurs the other way around, so that, say, a newly well-off agent persists in living in a poorer neighbourhood, it may be that she is willing to be an active proponent of social mixity and/or she is careful about what will become of her current income status in the future. Along with this social-mobility interpretation of type switching, let us remark that ethnic types may also be considered here, as these can change, be it through the agents' own feelings or the normative action of some externally imposed new definition.\footnote{The authors wish to thank an anonymous referee for drawing their attention to this particular point.}

For the sake of simplicity, we will henceforth refer to the switching dynamics as ``type mobility'', as opposed to the ``spatial mobility'' with preferential choice encoded in the standard Schelling dynamics. Let us however stress that the first interpretation (random allocation in an open system) should be equally kept in mind, in particular as it is compatible with open systems and points to a form of intervention that could be no less an actual mean of action in real urban contexts as the promotion of type mobility (especially if type mobility is to be social mobility in both directions!). Therefore, exploring random allocation strategies for housing sites can be viewed as an extension to the urban segregation problem of strategies for efficiency improvement through random noise that have gained attention lately, be it in relationship with democratic representation procedures~\cite{PGRSC}, hierarchical organizations~\cite{PRG2,PRG1}  or central bank interventions on financial markets~\cite{BPRH}.
\end{sloppypar}
\medskip

\begin{sloppypar}
We shall start from a most simple instance of a Schelling-class system, which meets the general criteria set out in~\cite{RM11}. A noticeable feature of the Schelling dynamics in our system is that the moves are ``blind'' ones, in the sense that when given the opportunity to move, an agent, if she does indeed move, will go to a site picked uniformly at random among vacant sites. Agents are therefore not satisfiers, let alone maximizers --- but this is not a pre-requisite for a system to belong to the general class defined in \cite{RM11}, as we understand it. The specifics of the algorithm implementing our model are detailed in section~\ref{sec:models+notations}. In section~\ref{sec:genres}, we comment on the general behaviour of the system and then derive in section~\ref{sec:PhDia}, from numerical simulations, cross-sections of the phase diagram. We compare these to the phase diagram in \cite{GVN09}, before focusing on the transitions that occur when the fraction of agents able to switch increases.
\end{sloppypar}

\section{System and notations}
\label{sec:models+notations}

Ours is a square-lattice system, featuring a finite number $width \times height$ of sites. At any instant, each site is in one of four states:
\begin{enumerate}
\item vacant,
\item occupied by an agent of (pure) type $A$,
\item occupied by an agent of (pure) type $B$,
\item occupied by an agent of type $C$,
\end{enumerate}
where $A$ and $B$ are the main two types, and $C$ corresponds to agents that have the ability to display either type~$A$ or type~$B$. There is a fixed number of agents of each three types (these numbers are determined by the occupation density $\rho$ and the fraction of $C$-agents $f$, given that there is an equal number of agents of pure type~$A$ and pure type~$B$).

We use periodic boundary conditions and discrete-time, ordered, asynchronous updating~\cite{Corn}. Agents are therefore examined in turn ; if they are of types $A$ and $B$, they follow a Schelling dynamics, \textit{viz.} they move to a vacant site (chosen uniformly at random) with probability $p_{h}$ if they are satisfied with their current position and with probability $p_{u}$ if they are not. The satisfaction of an agent of type $A$ or $B$ is either $0$ or $1$, depending on the proportion of their direct neighbours being of the opposite type: if this proportion is smaller than a given parameter $\tau$ (usually interpreted as the degree of ``tolerance'' of the agents~\cite{Schel1}) the agent is satisfied, otherwise she is not. Agents of type $C$ display at each instant one of the two types~$A$ or $B$, but do not move with the Schelling dynamics: when examined, they switch their displayed type with probability $p_s$, regardless of their neighbours' types. The number of $C$-type agents is expressed as a fraction $f$ of the total number of agents present in the system. Note that simulation code and results are available online.\footnote{~\url{https://sourceforge.net/p/phase-py/}}

\begin{table}
  \caption{ \label{tab:noise trans sim params}Simulation parameters}
  \centering
  \begin{tabular}{ll}
    \hline\noalign{\smallskip}
    Parameter & Value  \\
    \noalign{\smallskip}\hline\noalign{\smallskip}
    $\rho$ & $\in [0.9,1]$ \\
    $\tau$ & 0.3 \\
    $f$	& $\in [0,1]$\\
    $T_M$ & 500 \\
    $N_E$ & 50 \\
    $width, height$ & 30 \\
    $p_u$ & 0.2 \\
    $p_h$ & $10^{-4}$ \\
    $p_s$ & 0.05 \\
    \noalign{\smallskip}\hline
  \end{tabular}
\end{table}
Table~\ref{tab:noise trans sim params} summarizes the simulation parameters, with the typical values which they shall be given hereafter, unless otherwise stated. $\rho$, $\tau$, $f$, $p_h$, $p_u$ and $p_s$ were defined above. $T_M$ is the number of time steps  in a single realization (one time step corresponds to a complete round over all agents), and $N_E$ is the number of ensemble realizations. $width$ and $height$ are the dimensions in number of sites of the simulation grid. The values of~$\rho$ and~$\tau$ are chosen so that the pure Schelling dynamics ($f=0$) leads to a segregated state. (Note that in the basic Schelling-type dynamics we are using here, only direct neighbours are considered, which implies that values of $\tau$ tend to be distinguished only relative to the discrete values: $\frac{1}{4}$, $\frac{1}{2}$ and $\frac{3}{4}$; however, since vacant sites are not counted as neighbours, other values of $\tau$ can occasionally be distinguished.) The probability of moving for an agent unhappy with her current neighbourhood is set to a value significantly lower than $1$ to reflect difficulties (\textit{eg} economic factors) that may prevent unsatisfied agents from moving. The rate $p_s$ of switching for $C$-agents is then chosen to be four times smaller than $p_u$ to implement distinct timescales for the two types of dynamics (of course a reverse order of these timescales should also be studied later). Finally, the fact that agents who are statisfied with their current location are able to move ($p_h\neq 0$), even at a very small rate, suppresses certain simulation artifacts (see \textit{eg} \cite{StauSol}) and can also be interpreted as reflecting the possibility that agents make a ``wrong'' move due to misinformation or, again, economic factors.

As recalled in \cite{StauSol,GVN09,RM11}, Schelling-class systems are expected to reach some kind of equilibrium, or at least a steady state as far as segregation is concerned. This translates into the reaching of an almost constant limit value for a variable quantifying segregation within the system. Various choices of variables are possible. We shall follow~\cite{RM11} and work with a contact density $x(t)$, defined as the average over all occupied sites at time~$t$, of the ratio between the number of neighbours of the opposite type and the total number of neighbours (vacant neighbouring sites are not counted). The contact density is normalized by one half, so that when the two populations $A$ and $B$ are well mixed in the system, $x(t)$ is close to $1$, whereas it is close to $0$ when the system exhibits segregation. We shall write $\langle \dots \rangle$ for averages over all ensemble realizations. 

\section{General results}
\label{sec:genres}

\begin{figure}
  \resizebox{0.5\textwidth}{!}{
    \includegraphics{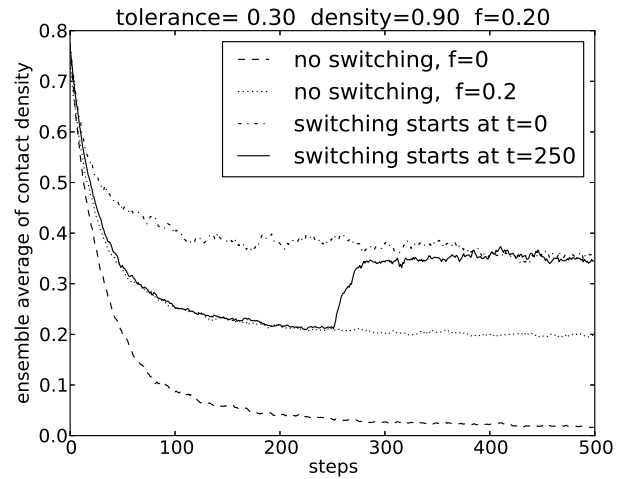}
  }
  \caption{Ensemble average $\langle x(t) \rangle$ of the contact density in four simulation modes.}
  \label{fig: dynam}
\end{figure}

\begin{sloppypar}
Let us first examine the influence of a non-zero fraction of switching agents on the limit value $x_\infty= \lim \langle x(t)\rangle$ of the contact density (time-averaged on the time-steps after a steady state has been reached), with four simulation modes: 
\begin{enumerate}
\item no switching agents are present ($f=0$);
\item switching agents are present ($f=0.2$) but do not switch ($p_s=0$);
\item a given fraction $f=0.2$ of switching agents participate, being ``activated'' only after a fixed delay ($250$~time steps);
\item a given fraction $f=0.2$ of switching agents participate, and are ``activated'' at the beginnning of the simulation.
\end{enumerate}
\end{sloppypar}

The first case is clear: this is a standard instance of a Schelling system.

For the second and third cases, note that $C$-type agents may be ``de-activated'', i.e. $p_s$ may be set to $0$ so that $C$-agents keep the same displayed type throughout the simulation. They thus act as fixed, non-moving agents of one or the other of the two types $A$ and~$B$.\footnote{Indeed in the second variant, $C$-agents stay at the same site and with the same displayed type for ever. Links with so called extremists in some opinion propagation models \cite{Def1} should be explored (the authors thank an anonymous referee for drawing their attention to this point).} Therefore, even though they are not themselves moving, they do participate in the dynamics by influencing the satisfaction of their neighbours, and by ``persisting'' in their residential choice whatever the current composition of their neighbourhood.

The fourth case simply corresponds to an activation of $C$-type agents from the beginning of the simulation.
\medskip

The occupation density is set to $0.9$, the tolerance $\tau$ to $0.3$. Starting from a well-mixed state (drawn from a uniform distribution), one observes that:
\begin{enumerate}
\item in the absence of switching agents, the system quickly relaxes to a segregated state with a value of $x_\infty$ very close to $0$, as expected for the chosen values of~$\rho$ and~$\tau$ (knowing the phase diagram of a typical Schelling system as in \cite{GVN09});
\item in the presence of switching agents that persist as fixed $A$ or $B$ agents, the relaxation is slower, and leads to a value of $x_\infty$ very close to $0.2$;
\item in the presence of switching agents that start as fixed $A$ or $B$ agents for $250$ time steps, relaxation starts similarly to the previous case; however, once activation takes place at $t=250$, the ensemble average $\langle x(t) \rangle$ makes a sudden jump to reach a higher limit value close to $0.35$;
\item in the full dynamics, i.e. with $C$-type agents activated from $t=0$, the system relaxes toward a moderate value of $x_\infty$ close to $0.35$.
\end{enumerate}

\begin{figure*}
  \centering
  \subfigure[Sample initial and final configurations in the absence of switching agents ($f=0$)]{
    \resizebox{0.58\textwidth}{!}{
      \includegraphics{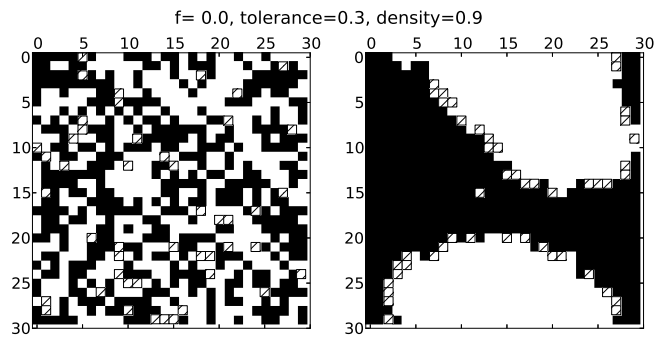}
    }
  }
  \centering
  \subfigure[Sample initial and final configurations in the presence of \textit{inactive} switching agents ($f=0.2$)]{
    \resizebox{0.58\textwidth}{!}{
      \includegraphics{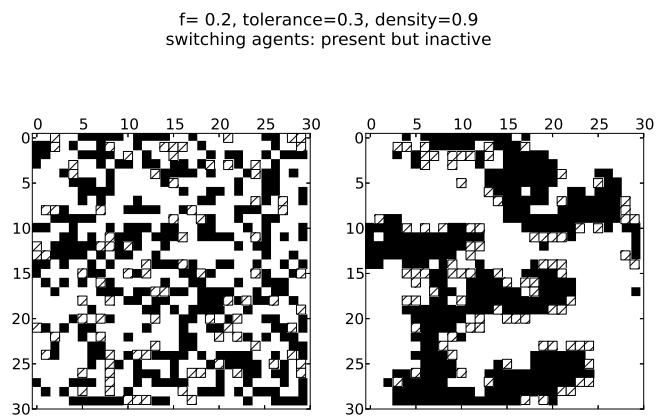}
    }
  }
  \centering
  \subfigure[Sample initial and final configurations in the presence of \textit{active} switching agents ($f=0.2$)]{
    \resizebox{0.58\textwidth}{!}{
      \includegraphics{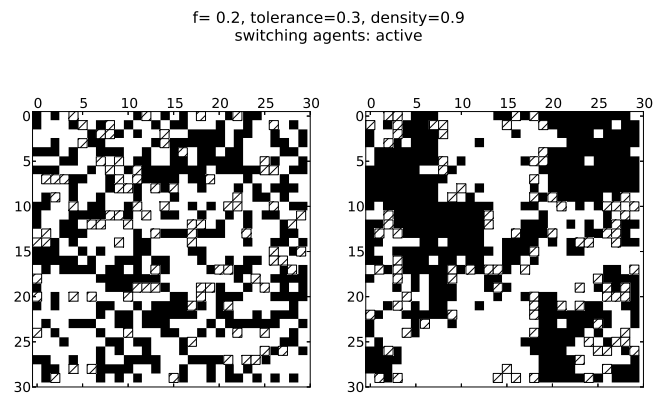}
    }
  }
  \caption{Sample initial and final configurations in the absence (a) or presence of active (c) or inactive (b) switching agents. Initial configurations are on the left-hand side. (Black~$=$~$A$~agents, white~$=$~$B$~agents, hatched~=~empty site, $C$~agents appear according to their current displayed type.)  Final configurations show strict phase separation in the absence of switching agents, a mixture of patches with fuzzy interfaces in the presence of active switching agents and an intermediate pattern in the presence of inactive switching agents.}
  \label{fig: states}
\end{figure*}

\begin{sloppypar}
The introduction of switching agents leads to a higher contact density, as could be intuitively expected. In the second variant, when $C$-agents are de-activated, the limit contact density $x_\infty$ is already significantly larger than in the pure Schelling variant. Actually it coincides with the value of $f$ in this particular case. This hints at the fact that it may simply be a background level of contact density due to local, point mixity around fixed, persisting $C$-agents that accounts for a higher value of $x_\infty$ --~rather than \textit{bona fide} mixity across the whole system. Visual observation of patterns in the final, steady state of our numerical simulations~(Fig.~\ref{fig: states}) shows that this type of effect is not the only thing happening in the presence of un-activated switching agents. Compared to the complete phase separation observed in the absence of switching agents, the final configurations observed in the presence of persisting agents is an intermediate step towards the more complex pattern (mixture of patches and fuzzy interfaces) formed in the presence of activated switching agents.\footnote{Note that in the presence of persisting agents, any form of phase separation is constrained by the fixed (and random) spatial pattern of persisting agents.}
\end{sloppypar}

The activation of switching, at time $t=250$ in the third variant and time $t=0$ in the fourth one, entices indeed a significant de-segregation phenomenon, leading quickly, in both cases, to a limit value $x_\infty \in [0.3,0.4]$. Such a value for the contact density can correspond to a mixture of phases, as described in~\cite{HaBe}: eg two ``pure'' clusters separated by a well-mixed area, or a mixture of homogeneous patches (see~Fig.~\ref{fig: states}).

Notice that such patchy mixtures are a way of ``optimizing'' the \textit{average} neighbourhood (but not necessarily the \textit{typical} neighbourhood) in a residential system at fixed $\tau < \frac{1}{2}$: indeed, a notable feature of Schelling systems is that, when they lie in the segregated phase, their average equilibrium contact density is significantly less than the agents' tolerance. This is precisely one of the features that made Schelling's model famous, since one could naively expect that when agents ``tolerate'' up to one half of neighbours to be of the type opposite to theirs, the system would in the long run reach some kind of steady state in which each agent had a neighbourhood comprising of one half of agents from the other type. But that was firstly assuming that \textit{average} neighbourhood and \textit{typical} neighbourhood would coincide, and secondly overlooking the possibility of a phase transition occuring at some value of the tolerance, especially a value smaller than one half ---~which is the case~\cite{Schel1,Schel2,Jones}.

\section{Phase diagram and transitions}
\label{sec:PhDia}

\label{sec:f0}

\begin{figure*}
  \centering
  \subfigure[Contact density]{
    \resizebox{18cm}{5cm}{
      \includegraphics{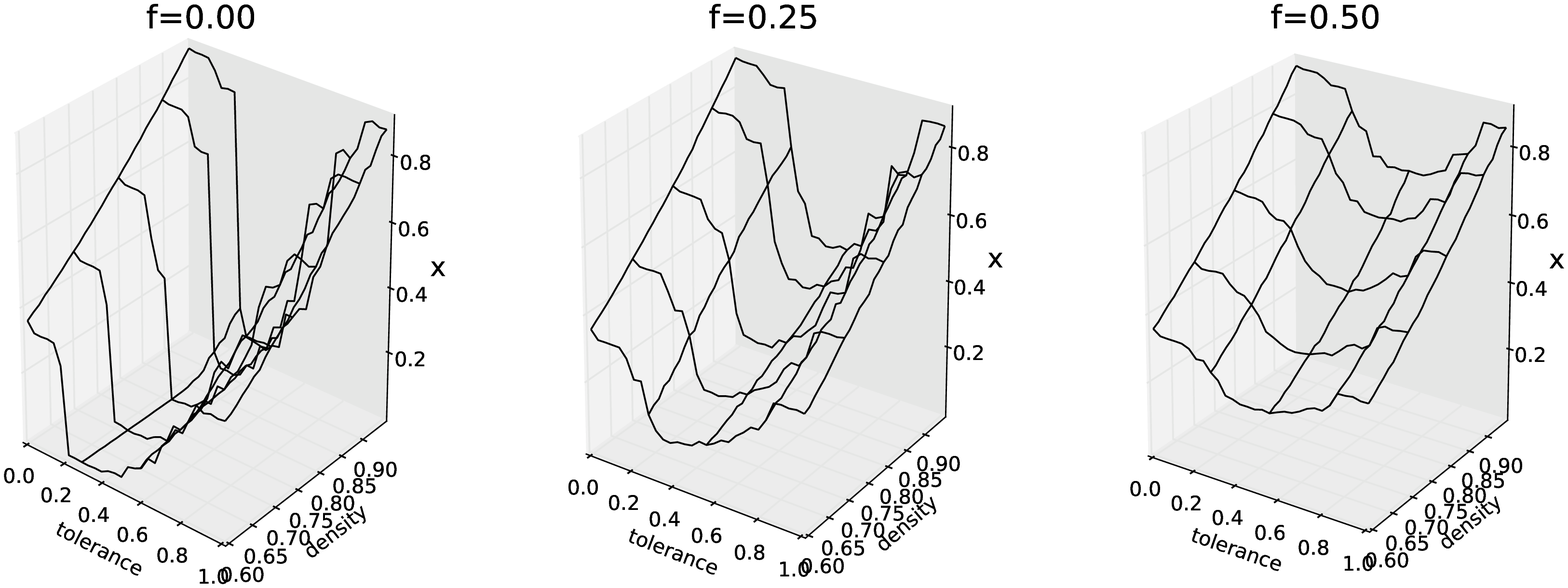}
    }
  }
  \centering
  \subfigure[Analogue of susceptibility (log of)]{
    \resizebox{18cm}{5cm}{
     \includegraphics{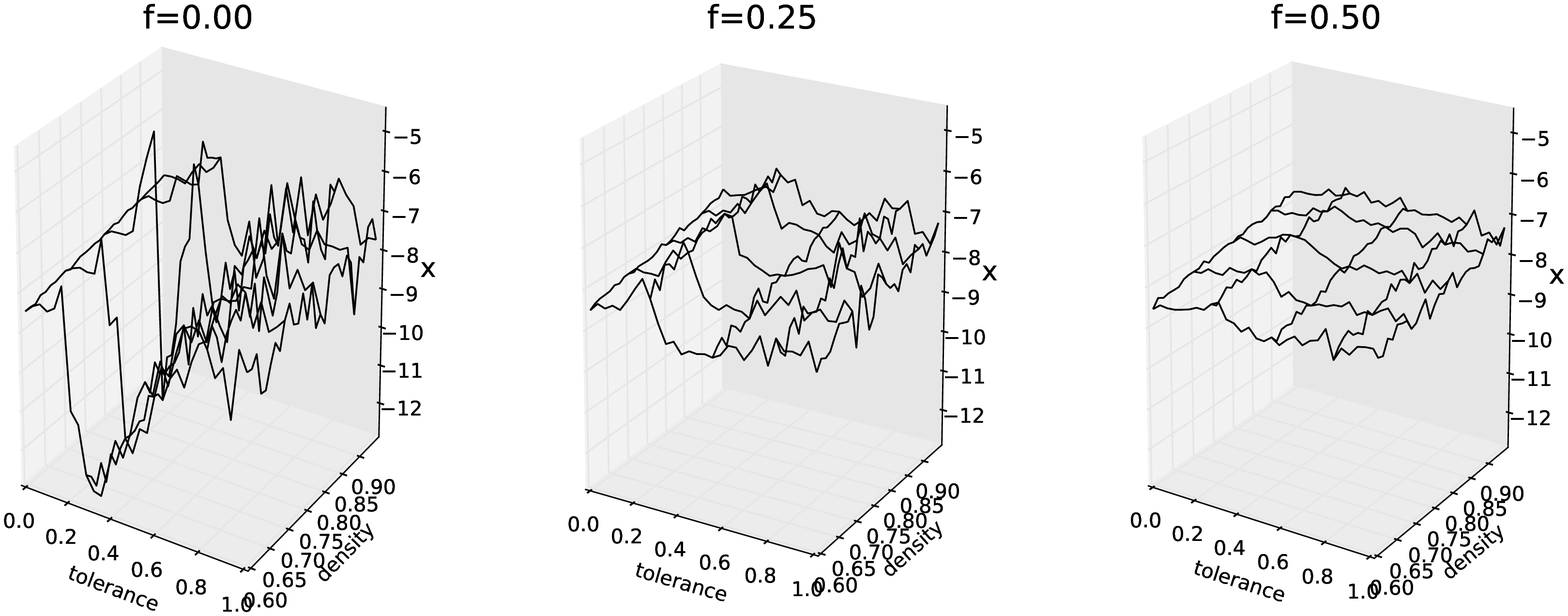}
    }
  }
  \caption{Contact density $x_\infty$ and analogue of susceptibility $\chi _\infty$, in terms of $\rho$ and $\tau$ as $f$ increases.}
  \label{fig: phase rhotol}
\end{figure*}

We wish now to look at the transitions occuring in the system both when $f=0$ and $\rho$, $\tau$ vary (where we should retrieve at least part of the results described in \cite{GVN09}), and when $f$ increases. Of course, our main interest lies in the latter case, and the relative bareness of our Schelling dynamics (with only direct neighbours taken into account and moves being blind) is not suited for an in-depth study of the system's phase diagram.

To facilitate further analysis of the system's behaviour and identify the transitions, we shall follow \cite{GVN09} and introduce, on top of the ensemble-averaged contact density $\langle x(t) \rangle$ and its limit $x_\infty$, a quantity analogous to the susceptibility of thermodynamical systems, and another analogous to the specific heat. The former is defined as $$\chi (t)=\frac{\langle [x(t)]^{2}\rangle - \langle x(t)\rangle ^{2}}{\tau},$$ and the latter as $$C(t)=\langle [E(t)]^{2}\rangle - \langle E(t)\rangle ^{2}, $$ where $E$ is defined in analogy with the energy in the Blume-Emery-Griffiths spin model \cite{BEG}: $$E=-\sum_{i,j}c_ic_j-(2\tau-1)\sum_{i,j}c_i^2c_j^2, $$ the sums being over pairs of neighbours, and $c_i=1$ for a site occupied by an agent of (displayed) type~$A$, $-1$ for an agent of (displayed) type~$B$ and $0$ for a vacant site. We shall write $\chi _\infty$ and $C_\infty$ for the time-averaged steady-state values of~$\chi(t)$ and~$C(t)$.

Gauvin \textit{et al.} have argued and explored the validity of such an analogy with thermodynamical spin systems~\cite{GVN09,GVNth}. We shall follow their lead in trying to identify transitions by looking for rapid changes in the value of $x_\infty$ accompanied by peaks in $\chi _\infty$ and/or $C_\infty$.
 
\subsection{Transitions at $f=0$ }
 
We look here briefly at the $(f=0)$~-section of the phase diagram, which corresponds to the first picture on the left in the top part of Fig.~\ref{fig: phase rhotol} for the $(f,\rho, \tau)$-diagram. Of course, our system is a much simplified version of the Schelling system studied in \cite{GVN09}, and in particular we expect changes mainly to occur at a couple of discrete values of~$\tau$:  $\frac{1}{4}$, $\frac{1}{2}$ and $\frac{3}{4}$. Fig.~\ref{fig: phase rhotol} shows that, at high occupation densities, our simulation results are indeed compatible with phase transitions occuring at $\tau = 0.25$ and $\tau = 0.75$. This is also supported by the behaviour of the susceptibility, as seen in the first figure on the left in the bottom part of Fig.~\ref{fig: phase rhotol}.

\begin{figure*}
  \centering
  \subfigure[Contact density]{
    \resizebox{12cm}{!}{
    \includegraphics{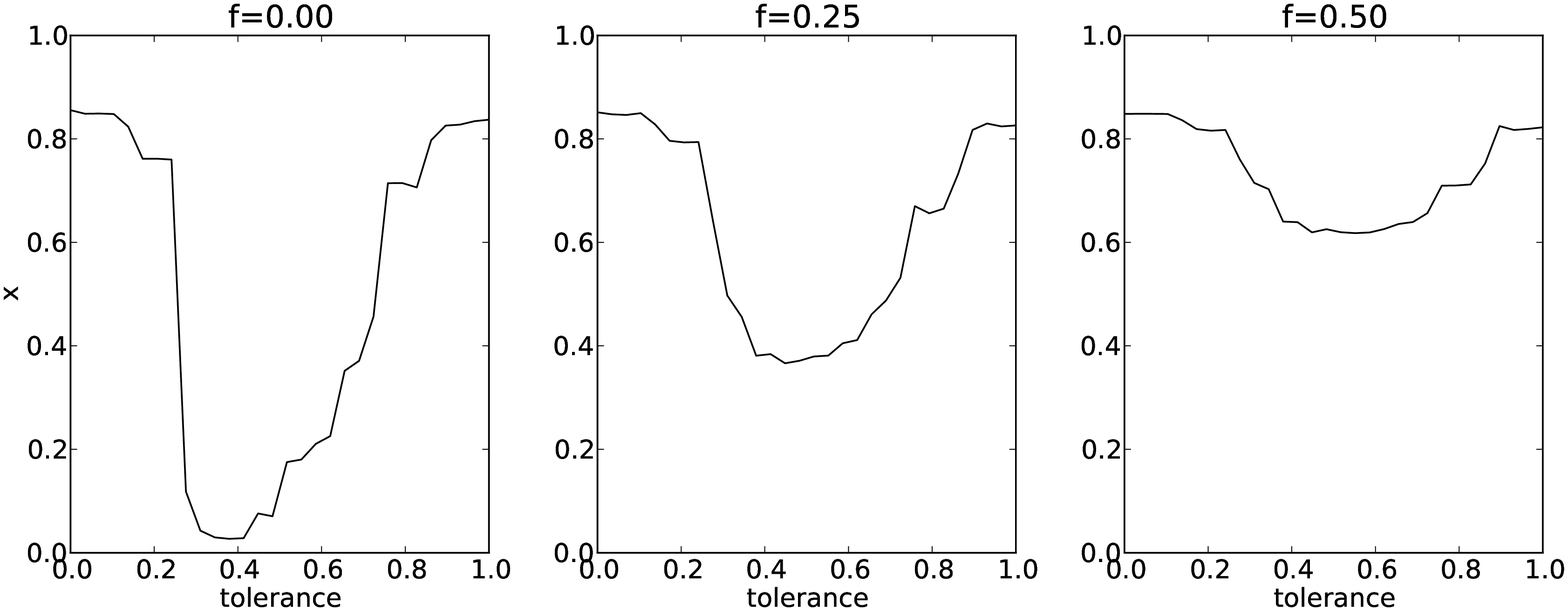}
    }
  }
  \centering
  \subfigure[Analogue of susceptibility (log of)]{
    \resizebox{12cm}{!}{
    \includegraphics{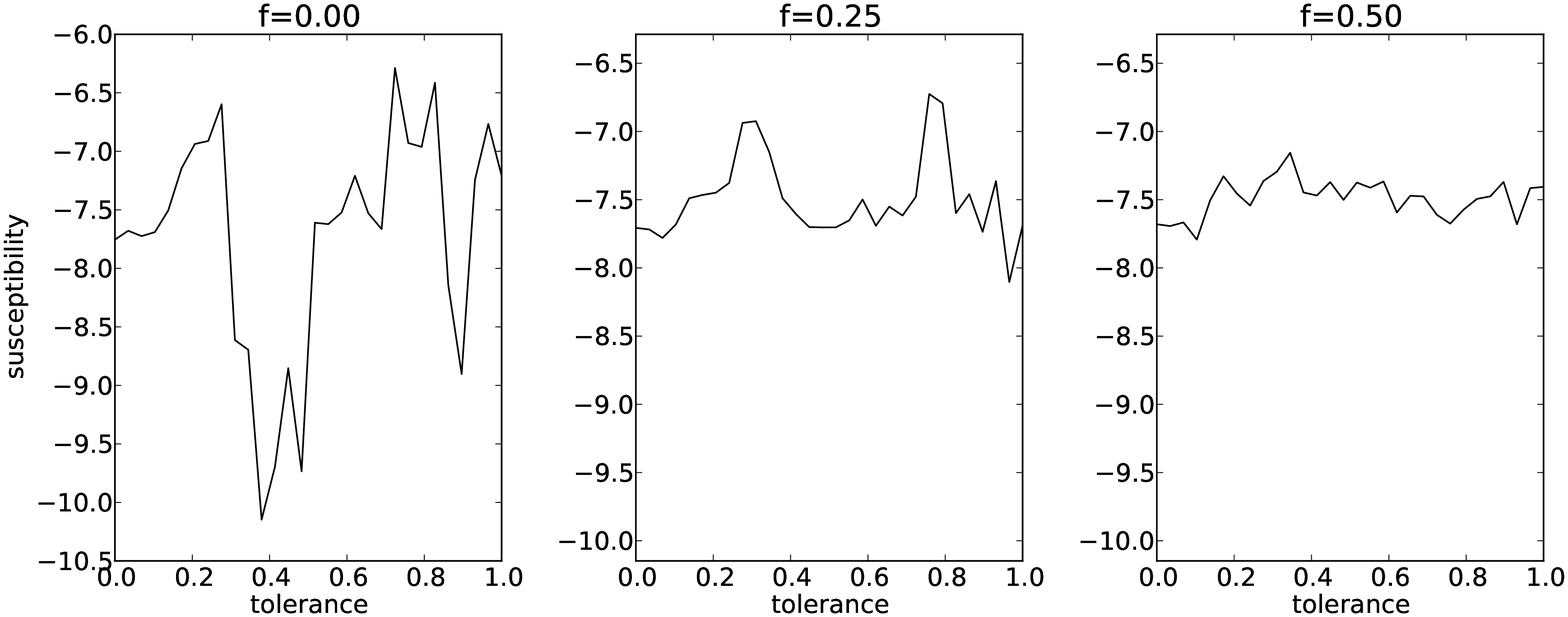}
    }
  }
  \caption{Profile of the contact density $x_\infty$ (a) and the analogue of the susceptibility $\chi _\infty$ (b) as $\tau$ varies, for a fixed occupation density $\rho=0.95$.}
  \label{fig: xprofile_fixedrho}
\end{figure*}

The change in the value of the contact density at $\tau = 0.25$ seems more abrupt than the one at $\tau = 0.75$, as indicated also by the profile of $x_\infty(\tau)$ at fixed $\rho=0.95$ (leftmost plot in Fig.~\ref{fig: xprofile_fixedrho}~(a)). Both transitions are marked by peaks of the susceptibility (leftmost plot in  Fig.~\ref{fig: xprofile_fixedrho}~(b)).  

In \cite{GVN09}, at low vacancy densities, the authors identified a first-order phase transition around $\tau_c = 0.75$ and a ``frozen'' state below $\tau _f = 0.4$. In between these values, their system would eventually reach a state of low contact density, that is, a segregated state. The transition from a state of frozen dynamics to the segregated phase was further investigated in \cite{RM12} and found to correspond to a jamming transition that could be reproduced \textit{via} a patch model and described by deterministic equations. In our system, the low-vacancy-density, low-tolerance state is separated from the segregated state by a more abrupt (possibly first-order) transition. This state can be understood as follows: at low $\tau$ (that is essentially $\tau=0$ as soon as $\tau < 0.25$), all agent are unsatisfied, unless they have only neighbours of the same type as theirs. Therefore, given the opportunity to move, they will, because moving does not depend on their finding a more ``welcoming'' site (contrary to what is the case in \cite{GVN09}, leading to the dynamics freezing). Eventually, blind moves lead to a reasonably well-mixed state, for reasons that differ from those behind the frozen, well-mixed state in \cite{GVN09} where the system simply remains in its initial state, which happens to be drawn from a well-mixed distribution.
\bigskip

Thus, the general shape of the phase diagram at\linebreak $f=0$ and high occupation density is very much comparable to that of a classical Schelling system as in \cite{GVN09}, and we proceed to the case~$f \neq 0$ in the next paragraph.

\subsection{Transitions along $f$}

\begin{sloppypar}
We examine now cross sections of the phase diagram at constant $\rho$, letting $f$, the fraction of switching agents present in the system, increase. Intuitively, if $f$ is sufficiently large, one expects a well-mixed system, as the situation then becomes that of fixed agents switching with equal probabilities to the A or B types. This is what one can observe in the upper part of Fig.~\ref{fig: (noise,tol) phasediag}, where indeed the mixed phase (high contact density) gains more and more ground as $f$ increases. As we have seen in section~\ref{sec:f0}, when $f=0$ two transitions are noticed, one near $\tau = 0.25$ and one near $\tau = 0.75$. Given that at large values of $f$, only a high contact-density phase survives, the corresponding transition lines in the $f$-$\tau$~plane should vanish or bend toward each other, jutting into the well-mixed phase and enclosing a segregated phase. This is precisely what one observes in the simulations, as pictured in the upper part of Fig.~\ref{fig: (noise,tol) phasediag}.
\end{sloppypar}

\begin{figure}
 \subfigure[Contact density]{ \resizebox{0.5\textwidth}{!}{
    \includegraphics{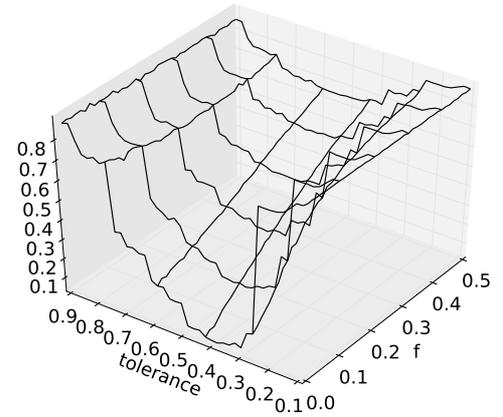}
  }
 }
 \subfigure[Analogue of susceptibility]{ \resizebox{0.5\textwidth}{!}{
    \includegraphics{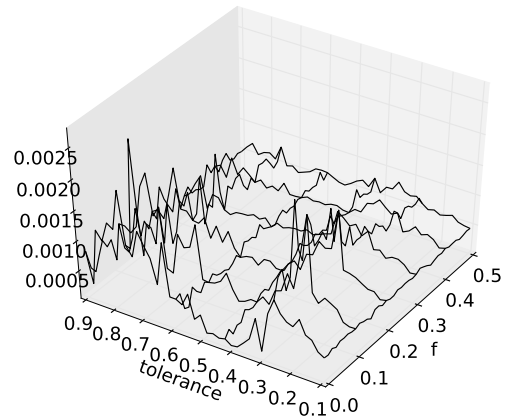}
  }
 }
 \subfigure[Analogue of specific heat]{ \resizebox{0.5\textwidth}{!}{
    \includegraphics{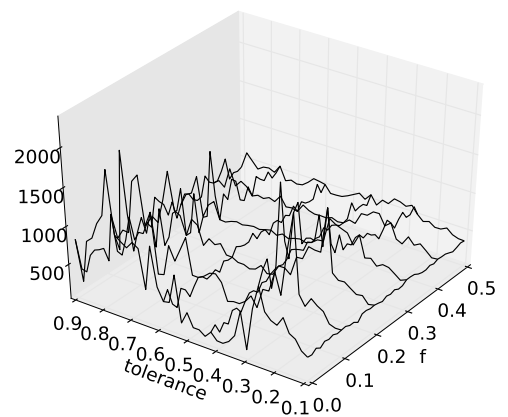}
  }
 }
  \caption{Phase diagram in terms of $f$ and $\tau$, at constant $\rho=0.95$. Contact density $x_\infty$ (a), analogue of susceptibility~$\chi _\infty$~(b), and analogue of specific heat~$C_\infty$~(c).}
  \label{fig: (noise,tol) phasediag}
\end{figure}

\begin{figure}
  \resizebox{0.5\textwidth}{!}{
    \includegraphics{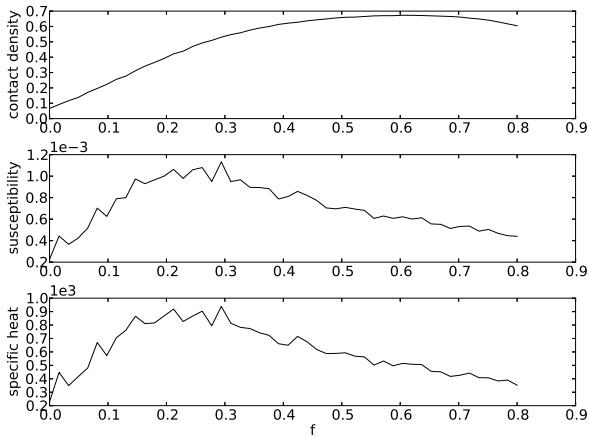}
  }
  \caption{Transition along $f$, at constant $\rho=0.95$ and $\tau=0.3$. Contact density $x_\infty$ and the analogues of susceptibility~$\chi _\infty$ and specific heat~$C _\infty$ are plotted as the fraction $f$ of switching agents increases.}
  \label{fig: noise trans}
\end{figure}

\begin{figure}
  \resizebox{0.45\textwidth}{!}{
    \includegraphics{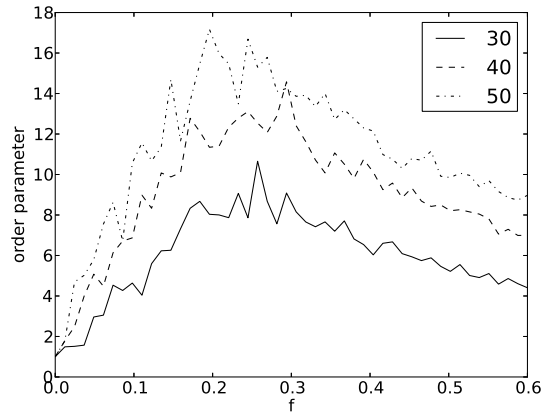}
  }
\caption{Finite-size effect. The analogue of specific heat $C_\infty$, normalized by its initial value, is plotted for various grid sizes, at $\rho =0.95$ and $\tau =0.3$ (with an ensemble of realizations other than the one used for~ Fig.~\ref{fig: noise trans}). The larger the system, the more marked is the peak, as one would expect in thermodynamical phase transitions.}
\label{fig: finite size}
\end{figure}

The nature of the transition occuring across these lines, at different points, is of particular interest, all the more as it varies with $f$: at small values of $f$, one has the transitions described in the previous subsection. They are compatible with  thermodynamical transitions, marked by peaks in the analogues of susceptibility and specific heat (lower parts of Fig.~\ref{fig: (noise,tol) phasediag}). Near higher values of $f$, the transition lines are no longer parallel to the $f$ axis, and thus can be crossed \textit{via} an increase in $f$. For instance, at fixed occupation density $\rho=0.95$ and tolerance $\tau=0.3$, Fig.~\ref{fig: noise trans} shows the profiles of the contact density, the analogue of susceptibility, and the analogue of specific heat as the fraction of switching agents $f$ is increased from $0$ to $1$. The change in the contact density is not abrupt but peaks can be observed in $\chi_\infty$ and $C_\infty$ around $f\approx 0.25$, with a finite-size effect \cite{Barb,Bin} shown in~Fig.~\ref{fig: finite size}.

Further up in the phase diagram shown in the top part of~Fig.~\ref{fig: (noise,tol) phasediag}, near $f\approx 0.5$, a more gradual change from a low to a high contact density is observed at intermediate values of the tolerance $\tau$ when $f$ is increased. No peaks are to be noted here in the susceptibility or specific heat analogues. This can indicate that the nature of the transition has changed along the line, from thermodynamical to simply a smooth, gradual change.

Coming back to the plot of $x_\infty$ against $f$ in~ Fig.~\ref{fig: noise trans}, an interesting feature lies in the fact that the contact density exhibits a maximum. If correct, this means that a better mixing (as measured by the contact density) could be achieved by tuning $f$ to some intermediate value, rather than simply setting it equal to $1$. In other words, a combination of Schelling dynamics (with segregative effects) and switching dynamics could bring more contact in the system than pure switching dynamics. However, this could be a bias in what the contact density actually measures or a bias linked to our sample of initial distributions: when $f=1$, where one would naively expect maximal mixing, the value of $x_\infty$ is liable to any pattern in the distribution of vacant sites since every agent remains fixed. Should it turn out to be an artefact linked to the initial distribution, confirmation of the existence of a maximum at some intermediate value of $f$ could still be interesting: one does have biased initial states in real cities. 

\section{Conclusion}

\begin{sloppypar}
We have examined the numerical behaviour of a system driven by a mixture of two dynamics: (i) a standard Schelling preference dynamics where agents of two types move stochastically across a grid according to their preferred composition of neighbourhood ---~which can be interpreted as residential choice, and (ii) a dynamics where some agents stay fixed but have the ability to switch from one type to the other, which can be interpreted either as ``type'' mobility or as random allocation for given sites (without the possibility that these remain vacant).

We have found that at $\rho=0.9$, $\tau=0.3$ and\linebreak $f=0.2$, the system does not reach full segregation, i.e. it does not exhibit eventually only two large, homogeneous domains, as pointed out already for other Schelling-class systems~\cite{Jones,StauSol}. A similar behaviour has been observed in a new implementation of our model, with parameter values set at~ $\rho=0.9$, $\tau=0.3$ and $f=0.15$~\cite{GHRF}. We will investigate this particular point in more detail by examining how the correlation length varies with~$f$, and how the type of contact (the nature of the interface) between $A$ and $B$ clusters changes.
\end{sloppypar}

Our findings also include the observation of a number of transitions as the control parameters $\rho$ (occupation density), $\tau$ (tolerance for neighbours of the opposite type) and $f$ (fraction of agents able to switch) are varied across their domains. Elements indicate in some cases thermodynamical transitions of the first order, in other cases continuous transitions, and in some instances dynamical transitions or simply gradual change. We will seek to confirm these with finer simulations, especially with a larger neighbourhood for each agent, paying particular attention to the presence of critical points. The exact nature of the final state in the various instances of the system should also be examined: is that an equilibrium state or a non equilibrium steady-state? Analytical descriptions of the system \cite{Grau1,Grau2}, at least in some cases, should help to better identify order and control parameters, and we will try to find mean field and/or simple analytical models that capture the dynamical onset of (de)segregation phenomena, in the manner of what was done in \cite{RM11} for standard Schelling systems.
\medskip

Another noticeable result of the simulations conducted for this article is the fact that, for some given, segregative values of the tolerance $\tau$ and occupation density $\rho$, the contact density shows a maximum along  a line of increasing fraction $f$ of switching agents. In other terms, without intervening on the occupation density nor on the agents' tolerance, mixed or at least lesser-segregated patterns of occupation can be achieved in the presence of a certain fraction of switching agents, without the need for that fraction to be close to $1$. This property will be further investigated, in particular the possibility of treating the presence of switching agents as noise in the system and observing noise-induced transitions \cite{GOS99}.
\medskip

\begin{sloppypar}
Finally, variants of the system introduced here will be explored in future work, beyond obvious simulations refinements on the Schelling part of the dynamics, with ``non-blind'' moves \cite{Schel2,Clark,LJ,Fos,GVN09}, random (a)synchronous updating, feedback loops (possibly local) acting on the tolerance of agents \cite{MSSt}, and extensions to more than two displayed types \cite{Ben,Schu,Fos}.

Indeed, if one focuses on the second interpretation of the model, it will be interesting to look at emphasizing distinct time-scales for what we have called type and spatial mobilities, and at asymmetrizing type mobility, \textit{eg} making it much more likely for $C$-agents displaying type~$B$ to switch to type~$A$ rather than for $C$-agents displaying type~$A$ to switch to type~$B$, as might be the case in real socio-economic contexts; or allowing $C$-type agents to move on the grid, according to their displayed type, in an asymmetric manner \textit{eg} by allowing $C$-agents to move when displaying type~$A$ but not when displaying type~$B$.

Exploration of these numerous variants will find useful leads in the use of real data \cite{Ben2,BrM,Fos,KS,HaBe} ; specifically, with respect to the first interpretation of the model (landlords and/or a regulatory agency implementing, on some fixed housing sites, an allocation policy that is blind to agent types) we will seek data from housing associations. We will also pursue the development of interdisciplinary work with sociologists, geographers and urban-system scientists (\cite{Stau} rightly criticizes the lack of interdisciplinary work in the literature related to the Schelling model).
\end{sloppypar}

\begin{acknowledgement}
The authors wish to thank two anonymous referees for their careful reading of the manuscript and their helpful comments and remarks.
\end{acknowledgement}

\bibliographystyle{Science}
\bibliography{schelling2}

\end{document}